\documentclass{elsart}
\usepackage{graphics,natbib,graphicx,psfig,amssymb,ccaption}
\usepackage{lscape} 
\journal{New Astronomy}
\input psfig.sty
\begin{document}
\begin{frontmatter}
\title{First identification and absolute magnitudes of the red clump stars 
in the Solar neighbourhood for {\em WISE}}
\author[istanbul]{E. Yaz G\"okce\corauthref{cor}},
\corauth[cor]{Corresponding author. Fax: +90 212 440 03 70}
\ead{esmayaz@istanbul.edu.tr}
\author[istanbul]{S. Bilir},
\author[istanbul]{N. D. \"Ozt\"urkmen},
\author[istanbul]{\c S. Duran},
\author[istanbul]{T. Ak},
\author[istanbul]{S. Ak},
\author[istanbul]{S. Karaali}
\address[istanbul]{Istanbul University, Faculty of Science, Department of Astronomy 
and Space Sciences, 34119 University, Istanbul, Turkey}

\begin{abstract}
We present the first determination of absolute magnitudes for the red clump (RC) 
stars with the Wide--field Infrared Survey Explorer {\em (WISE)}. We used recently
reduced parallaxes taken from the {\em Hipparcos} catalogue and identified 3889 RC stars 
with the {\em WISE} photometry in the Solar neighbourhood. Mode values estimated from 
the distributions of absolute magnitudes and a colour of the RC stars in {\em WISE} photometry are 
$M_{W1}=-1.635\pm 0.026$, $M_{W3}=-1.606\pm 0.024$ and $(W1-W3)_0=-0.028 \pm 0.001$ mag. 
These values are consistent with those obtained from the transformation formulae using 
2MASS data. Distances of the RC stars estimated by using their $M_{W1}$ and $M_{W3}$ absolute 
magnitudes are in agreement with the ones calculated by the spectrophotometric method, 
as well. These {\em WISE} absolute magnitudes can be used in astrophysical 
researches where distance plays an important role. 
\end{abstract}

\begin{keyword}
97.10.Vm Distances, parallaxes  
\sep 97.20.Li Giant and subgiant stars  
\sep 98.35.Pr Solar neighbourhood   
\end{keyword}

\end{frontmatter}

\section{Introduction}

The RC stars are core-helium-burning stars on the horizontal branch of the Hertzsprung--Russell 
diagram. They lie on the red counterpart of the horizontal branch in the colour-magnitude 
diagrams of the globular clusters \citep{Can70,Fau73}. One of the most important features of the 
RC stars is their very limited luminosity distribution in all photometric systems which makes 
them standard candles for estimating astronomical distances in Galactic and extra--galactic 
scales. Thus, a mean absolute magnitude obtained for the RC stars in any photometric system may 
be very useful in studies where distance is a crucial parameter. 

Using {\em Hipparcos} astrometric and photometric data \citep{ESA97}, \citet{Keenan99} 
found the spectral types (G8III-K2III) of the RC stars in the Solar neighbourhood and estimated 
that their absolute magnitudes are $0.7\leq M_V \leq 1.0$ mag. \citet{Al00}, who combined the 
near-infrared photometry (NIR) and {\em Hipparcos} parallax data \citep{ESA97}, obtained that the 
mean absolute magnitude of the RC stars is $M_{K_{s}}=-1.61\pm0.03$ mag in the $K_{s}$-band. With 
the newly reduced {\em Hipparcos} parallaxes \citep{vanLee07}, \citet{Groenewegen08} estimated a 
mean $K_{s}$-band absolute magnitude of $M_{K_{s}}=-1.54\pm0.04$ mag for the RC stars with a 
negligible dependence on the metallicity. By re-observing the RC stars in the sample of 
\citet{Al00}, \citet*{Laney12} obtained a mean absolute magnitude of $M_{K_{s}}=-1.613\pm0.015$ 
mag for these stars, which is a very consistent value with the mean in \citet{Al00}. In one of 
the most recent studies, \citet{Bilir13a} calibrated the $M_{K_{s}}$, $M_V$, $M_J$ and $M_g$ 
absolute magnitudes of the RC stars in terms of the colours in the Johnson--Cousins, Two Micron 
All Sky Survey \citep[2MASS,][]{Skrutskie06} and the Sloan Digital Sky Survey 
\citep[SDSS,][]{York00} photometric systems. 

{\em WISE} is the recent generation infrared (IR) space telescope with much higher sensitivity 
than previous IR survey missions \citep{Wright10}. It began surveying the sky on 2010, January 14 
and completed its full coverage of the sky on 2010, July 17. The all sky data were released on 
2012, March 14. {\em WISE} surveyed the entire sky in four mid--infrared (MIR) bands, i.e. 3.4, 4.6, 12 
and 22 $\mu$m. These bands are denoted as $W1$, $W2$, $W3$ and $W4$ with the angular resolutions 
6.1, 6.4, 6.5 and 12 arcsec, respectively. {\em WISE} has achieved 5$\sigma$ point source 
sensitivities better than 0.08, 0.11, 1 and 6 mJy at the $W1$, $W2$, $W3$ and $W4$ bands, 
respectively. These sensitivities correspond to the Vega magnitudes of 16.5, 15.5, 11.2 and 7.9 mag. 
Thus, {\em WISE} will go one magnitude deeper than the 2MASS $K_{s}$-band data in $W1$ for sources 
with spectra close to that of an A0 star and even deeper for moderately red sources like K stars 
or galaxies with old stellar populations. The passband profiles for the 2MASS \citep{Cohen03} 
and {\em WISE} \citep{Wright10} photometric systems are shown in Fig. 1. {\em WISE} is the 
ideal photometric system for investigating RC stars. It is almost hundred times more sensitive 
than other infrared surveys, e.g. {\em IRAS} \citep[InfraRed Astronomical Satellite,][]{Neuge84}, 
{\em COBE} \citep[COsmic Background Explorer,][]{Smoot92} and {\em AKARI} \citep{Mura04}.

The mean absolute magnitudes of the RC stars can be affected by Galactic population differences. 
A weak dependence of the $I$-band absolute magnitude of the RC stars in the Solar neighbourhood 
on $[Fe/H]$ was found by \citet{Udalski00}. He suggested an absolute magnitude 
correction of 0.07 mag in the $I$-band for Large Magellanic Cloud. \citet{Girardi01} and 
\citet{Salaris02} predicted mean absolute magnitudes in $V$, $I$ and $K$ bands for the RC stars 
for a large range of ages and metallicities from the models of \citet{Girardi00b}. According to 
their models, the absolute magnitudes of the RC stars in $I$-band depends both on metallicity and age.
Metallicity dependence of $V$-band absolute magnitudes for the RC stars was recently studied by 
\cite{Bilir13b} using a sample of globular and open clusters. They presented 
an absolute magnitude calibration including the $B-V$ colour and $[Fe/H]$ metallicity of the RC 
stars in the clusters with a wide range of metallicities. As mentioned above, 
\citet{Groenewegen08} suggested that dependence of the mean $K_{s}$-band absolute magnitude of 
the RC stars on metallicity is negligible. Although there are neither theoretical nor observational 
studies for the metallicity dependence of the {\em WISE} photometric bands on age and/or metallicity, 
it may be estimated that this dependence, if present, must not be important.

As {\em WISE} can go at least one magnitude deeper than the 2MASS for the RC stars, a mean 
absolute magnitude of these stars in any {\em WISE} photometric band may be very useful. Thus, 
their mean magnitudes and locations in the colour spaces should be estimated using newly released 
{\em WISE} data \citep{Cutri12}. In this study, we aim to estimate mean absolute magnitudes for 
the RC stars with MIR photometry, i.e. $W1$ and $W3$. In section 2 we present the data. Identification 
of the RC stars in {\em WISE} is given in Section 3. Section 4 is devoted to testing the absolute 
magnitude estimation, and finally, we give conclusions in Section 5.

\section{The data}

As the aim of this study is to estimate MIR absolute magnitudes of the RC stars using their precise 
trigonometric parallaxes, the RC stars in the Solar neighbourhood should be selected. Thus, the 
main sample in this study was constructed using the newly reduced {\em Hipparcos} catalogue 
\citep{vanLee07}. Stars with the relative parallax errors $0 < \sigma_\pi / \pi \leq 0.15$ were 
selected from the {\em Hipparcos} catalogue which allows us to study 
a stellar sample with precise distances in the Solar neighbourhood. In order to obtain 
2MASS and {\em WISE} magnitudes of the stars selected from the {\em Hipparcos} catalogue, we 
matched the selected sample with the available 2MASS Point Source Catalog \citep{Cutri03} and 
{\em WISE} All-Sky Data Release \citep{Cutri12} according to the equatorial coordinates. The number 
of stars matched in the both photometric system is 46,214 which are from different luminosity classes 
and spectral types. 

The observed trigonometric parallaxes are systematically biased, called Lutz-Kelker 
bias \citep{LK73}, and this should be corrected for the stars with $0 < \sigma_\pi / \pi \leq 0.175$. 
Thus, we corrected the observed {\em Hipparcos} parallaxes of the stars in our sample using the 
method in \citet{Smith87} who described the Lutz-Kelker correction (LK) analytically, as follows

\begin{eqnarray}
\pi_0=\pi \left(\frac{1}{2} + \frac{1}{2} ~ \sqrt[]{1-16(\sigma_\pi / \pi)^2}\right),
\end{eqnarray}
where $\pi$ and $\pi_0$ are the observed and corrected parallaxes, respectively, and $\sigma_\pi$ 
denotes the observed parallax error. 

Although we study the stars in the Solar neighbourhood, the reddening should be taken into 
account. Thus, we used the dust maps taken from \cite{Schlegel98} and evaluated an $E_{\infty}(B-V)$ 
colour excess for each star. We then reduced them using the procedure described by 
\cite{Bahcall80}. Details of the procedure can be found in \cite{Bilir13a}. Distributions of the 
reduced colour excesses of the stars in our sample (46,214 stars) according to the Galactic 
latitude are shown in Fig. 2. The median value of the reduced colour excesses for all the Galactic 
latitude intervals is $\sim$0.03 mag.

\section{Identification of the RC stars in {\em WISE}}

\citet{Bilir12} showed that the RC stars can be effectively detectable in $W1$, $W2$ and $W3$ bands 
of the {\em WISE} photometry while dwarfs can be found using $W1$ and $W2$ \citep{Bilir11}. We preferred 
$W1$ and $W3$ bands in order to identify the RC stars in {\em WISE}. For de--reddening the magnitudes 
$W1$ and $W3$, we adopted the equations in \cite{Bilir11} derived by means of a spline function fitted 
to the data of \citet{Card89} which cover a range of $0.002 \leq \lambda \leq 250$ $\mu$m:

\begin{eqnarray}
(W1)_0=W1-0.158 \times E(B-V),\\\nonumber
(W3)_0=W3-0.087 \times E(B-V).\\\nonumber
\end{eqnarray}

Absolute magnitudes were calculated by setting the {\em Hipparcos} parallaxes, apparent magnitudes and 
total absorptions in the Pogson's equation. We constructed the $M_{W1}$-$(W1-W3)_0$ colour-magnitude 
diagram for the 46,214 sample stars obtained from the {\em Hipparcos} data as explained in Sect. 2. 
The $M_{W1}$-$(W1-W3)_0$ colour-magnitude diagram of these stars is shown in Fig. 3. The RC stars 
can be prominently detectable by an eye inspection in a region defined by a rectangle constrained 
with $-2.35 \leq M_{W1} \leq -1.00$ and $-0.10 < (W1-W3)_{0} < +0.05$ mag. The central position of 
this adopted region is roughly ($M_{W1}$, $(W1-W3)_{0}$) = (-1.64, -0.03) mag and there are 3889 
stars in it.

Median $E(B-V)$ colour excess for these most probable RC stars is 0.031 mag. Distribution of the relative 
parallax errors of the stars is shown in Fig. 4 with a median value of 0.1. {\em WISE} photometric errors of 
the RC stars are shown in Fig. 5. Photometric errors for the stars brighter than 4 mag are roughly between 
0.1-0.2 mag in the $W1$ filter, while they are smaller for fainter stars. Errors for the $W3$ filter are 
almost the same for all stars in the sample. Median errors for the $W1$ and $W3$ filters are 0.081 and 
0.015 mag, respectively. Median colour error for the RC stars is $(W1-W3)_{err}=0.082$ mag. 

In order to test the dependence of the mean absolute magnitudes in $WISE$-bands on the relative parallax 
errors, we created three sub-samples of the RC stars according to the relative parallax errors of 
$\sigma_\pi / \pi \leq 0.05$, $\sigma_\pi / \pi \leq 0.10$ and $\sigma_\pi / \pi \leq 0.15$. 
Distributions of the $M_{W1}$ and $M_{W3}$ absolute magnitudes and $(W1-W3)_0$ colour for the RC stars 
in these sub-samples are shown in Fig. 6. Gaussian fits applied to the distributions are also shown.
Modes of the absolute magnitudes and a colour in {\em WISE} obtained by fitting the Gaussian functions to 
the corresponding histograms in Fig. 6 are listed in Table 1. 
As can be seen from Table 1, the LK-corrected mean absolute magnitudes and the colour are decreasing 
with increasing relative parallax errors. Although the parallax errors are very small for the stars in 
the second column of Table 1, the number of stars with $\sigma_\pi / \pi \leq 0.05$ is about ten times 
smaller than the ones with $\sigma_\pi / \pi \leq 0.15$. While the number of stars increases about ten 
times, the mean absolute magnitudes and colour decrease about 0.06 and 0.007 mag, respectively. This small 
change indicates the reliability of the sample in this study. The mean absolute magnitudes and colors 
for the sub-samples in the last two columns of Table 1 are in agreement. Although the LK-correction for 
the stars with $\sigma_\pi / \pi \geq 0.10$ is larger, this agreement demonstrates that there are no 
crucial LK-correction effects on the mean values obtained for the sub-samples in the last two columns of 
Table 1. Moreover, the number of stars with $\sigma_\pi / \pi \leq 0.15$ is about two times larger than 
the ones with $\sigma_\pi / \pi \leq 0.10$. With these reasons, we adopted the mean values obtained for 
the stars with $\sigma_\pi / \pi \leq 0.15$ in this study.

\begin{table}
\setlength{\tabcolsep}{6pt}
\renewcommand{\arraystretch}{1}
\center 
\caption{Mode values of the Gaussian fits applied to the absolute magnitude and colour distributions 
in {\em WISE} for three different relative parallax error limits. $N$ represents the number of stars.} 
\begin{tabular}{lccc}
\hline
               & $\sigma_{\pi}/\pi \leq 0.05$  & $\sigma_{\pi}/\pi \leq 0.10$ & $\sigma_{\pi}/\pi \leq 0.15$  \\
\hline
  $M_{W1}$     & $ -1.576\pm 0.024$ & $ -1.612\pm 0.022$ & $ -1.635\pm 0.026$ \\
  $M_{W3}$     & $ -1.552\pm 0.020$ & $ -1.585\pm 0.019$ & $ -1.606\pm 0.024$ \\
  $(W1-W3)_0$  & $ -0.021\pm 0.002$ & $ -0.026\pm 0.001$ & $ -0.028\pm 0.001$ \\
  $N$          & 397                & 1969               & 3889 \\
\hline
\end{tabular}  
\end{table}

\section{Testing the {\em WISE} absolute magnitudes and colour of the RC stars}

\subsection{Absolute magnitudes from transformation equations}

Absolute magnitudes and colour of the RC stars estimated in the previous section for {\em WISE} can be 
tested using known transformation equations that use magnitudes and colours of the 2MASS photometric system. 
In order to use the transformation equations, we should identify the 
RC stars in a colour-magnitude diagram of the 2MASS photometric system. By the procedure explained 
in Sect. 2 for the {\em WISE} data, we de-reddened the 2MASS photometric data of 46,214 stars selected 
from the {\em Hipparcos} catalogue. The 2MASS colours and magnitudes of the stars were de-reddened 
using the reduced colour excesses and the following equations from \citet{FioM03} and \citet{Bilir08} for 
the 2MASS photometric system,

\begin{eqnarray}
J_{o}=J-0.887\times E(B-V), \nonumber \\
(J-H)_{o}= (J-H)-0.322\times E(B-V),\nonumber \\
(H-K_{s})_{o}=(H-K_{s})-0.183\times E(B-V),\\ \nonumber
(J-K_{s})_{o}=(J-K_{s})-0.505\times E(B-V).\nonumber
\end{eqnarray}

We used empirical colour--magnitude diagrams to determine the locations of the RC stars. Firstly, we plotted 
the $M_{K_{s}}$ absolute magnitudes versus $(J-K_{s})_0$ colours of 46,214 stars in Fig. 7 and identified 
the RC stars with two constraints, i.e. $-2.25 \leq M_{K_{s}} \leq -1.00$ and 
$0.50 \leq (J-K_{s})_0 \leq 0.75$ mag, by an eye inspection. With this procedure, we identified the most 
probable 2937 RC stars in our sample. Median colour excess for these stars is 0.026 mag which is 
in agreement with the median value calculated for the whole sample including 46,214 stars. Distribution 
of the relative parallax errors of the 2937 RC stars is shown in Fig. 8. The median value of the 
relative parallax errors is 0.097. 

2MASS photometric errors of these most probable RC stars are shown in Fig. 9 that presents two different 
error populations in the $J$, $H$ and $K_s$ magnitude measurements. These populations originates from the 
saturation of bright stars with $J \lesssim 5$ mag, about 32 per cent of the sample. This analysis shows that 
signal-to-noise ratio is greater than 20 ($J_{err} \lesssim 0.1$ mag) for most of the stars in the sample. 
In this case, median magnitude errors for the 2937 RC stars are $J_{err}=0.027$, 
$H_{err}=0.042$ and $K_{s_{err}}=0.021$ mag. Two different populations are also seen in the distribution 
of NIR colours. Median colour errors for the sample are $(J-H)_{err}=0.050$, $(H-K_{s})_{err}=0.047$ 
and $(J-K_{s})_{err}=0.034$ mag. Distributions of the $M_{J}$, $M_{H}$ and $M_{K_{s}}$ absolute 
magnitudes and the $(J-H)_0$, $(H-K_{s})_0$ and $(J-K_{s})_0$ colours of the 2937 RC stars are shown in Fig. 10. 
Gaussian fits applied to the distributions are also shown in the same figure. Modes of the absolute 
magnitudes and colours in 2MASS obtained by fitting the Gaussian functions 
to the corresponding histograms in Fig. 10 are listed in Table 2. Internal errors for the modes of 
the absolute magnitude and colour distributions are $M_{J_{err}}=0.016$, 
$M_{H_{err}}=0.014$, $M_{K_{s_{err}}}=0.025$, $(J-H)_{err}=0.002$, $(H-K_{s})_{err}=0.002$ and 
$(J-K_{s})_{err}=0.003$ mag.

The $K_{s}$ absolute magnitude in Table 2 is smaller than the value given by \citet{Groenewegen08} 
($M_{K_{s}}=-1.54\pm0.04$ mag) while it is greater than the values in \citet{Al00} 
($M_{K_{s}}=-1.61\pm0.03$ mag) and \citet{Laney12} ($M_{K_{s}}=-1.613\pm0.015$ mag). \citet{Laney12} 
observed 226 RC stars in the Solar neighbourhood in near-infrared $JHK$-bands  and calculated mean 
absolute magnitudes of the RC stars using {\em Hipparcos} parallaxes. The mean absolute magnitudes 
given by \citet{Laney12} are $M_{J}=-0.984\pm0.014$, $M_{H}=-1.490\pm0.015$ and 
$M_{K_{s}}=-1.613\pm0.015$ mag. Differences between the mode values in this study (Table 2) and the 
ones in \citet{Laney12} are not larger than 0.030 mag. Using these comparisons, we conclude that 
the mean absolute magitudes and colours of the RC stars in our study are in good agreement with the 
ones in the previous studies, although the saturated stars mentioned above are included in the 
calculations.

\begin{table}
\setlength{\tabcolsep}{2pt}
\center 
\caption{Modes of the absolute magnitudes, colours and internal errors in 2MASS obtained 
by fitting the Gaussian functions to the corresponding histograms in Fig. 10.} 
\begin{tabular}{lc}
\hline
                          & Mode (mag) \\
\hline
  $M_J$         & $-0.970 \pm 0.016$ \\
  $M_H$         & $-1.462 \pm 0.014$ \\  
  $M_{K_{s}}$   & $-1.595 \pm 0.025$ \\
  $(J-H)_0$     & $ ~~0.485 \pm 0.002$ \\ 
  $(H-K_{s})_0$ & $ ~~0.130 \pm 0.002$ \\
  $(J-K_{s})_0$ & $ ~~0.612 \pm 0.003$ \\
\hline
\end{tabular}
\end{table}

We tested the adopted absolute magnitudes and colours of the RC stars estimated with 
the {\em WISE} data using equations evaluated by \citet{Bilir12} which transform magnitudes 
and colours from the 2MASS photometric system to the {\em WISE} photometric system and vice 
versa. Thus, the $M_{W1}$, $M_{W3}$ absolute magnitudes and $(W1-W3)_0$ colour were evaluated 
using these transformation equations and compared them with the corresponding ones estimated 
in this study. As the metallicities of the RC stars in the sample are not known and metallicity 
dependence of {\em WISE} magnitudes and colours of these stars must be very weak, we preferred 
the metal free transformation equations as follows \citep{Bilir12}:

\begin{eqnarray}
(J-W1)_{0} = 1.102(55) (J-H)_{0} + 0.737(72) (H-K_{s})_{0} + 0.035(31), \nonumber \\
(J-W3)_{0} = 1.029(64) (J-H)_{0} + 0.780(84) (H-K_{s})_{0} + 0.106(36).\\ \nonumber
\end{eqnarray}

Here the numbers in parenthesis are the errors of coefficients for the last two digits. By setting 
the 2MASS magnitudes and colours in Table 2 in the transformation equations, we obtained the 
semi-empirical {\em WISE} $M_{W1}$, $M_{W3}$ absolute magnitudes and $(W1-W3)_0$ colour and listed 
them in Table 3. In this table, the adopted absolute magnitudes and a colour in {\em WISE} (Table 1) and 
the evaluated ones from the transformation equations are compared.

\begin{table}
\setlength{\tabcolsep}{4pt}
\renewcommand{\arraystretch}{1}
\center 
\caption{Modes of distributions for the absolute magnitudes and a colour in {\em WISE} for the RC stars 
with $\sigma_\pi / \pi \leq 0.15$ (see Table 1). The evaluated values from the transformation 
equations \citep[TEs;][]{Bilir12} are also given. Differences between two sets of the data are 
in the last column.} 
\begin{tabular}{lccc}
\hline
              &      Mode (mag)   & Value from TEs (mag) & Difference (mag) \\
\hline
  $M_{W1}$    & $ -1.635\pm 0.026$ & $ -1.635\pm 0.024 $ & $ ~~0.000 $ \\
  $M_{W3}$    & $ -1.606\pm 0.024$ & $ -1.676\pm 0.028 $ & $ +0.070 $ \\
  $(W1-W3)_0$ & $ -0.028\pm 0.001$ & $ +0.041\pm 0.037 $ & $ -0.069 $ \\
\hline
\end{tabular}  
\end{table}

Table 3 demonstrates that the {\em WISE} absolute magnitudes and colour estimated for the RC stars 
in this study are in very good agreement with the semi-empirical ones evaluated from the transformation 
equations within the error limits. While $W1$ is the same as evaluated from the transformation 
equations, the original $W3$ magnitude deviates from the evaluted one within the error limits. 
Difference between the estimated and evaluated $(W1-W3)_0$ colours originates from the deviation 
of $W3$ magnitude, as expected.

\subsection{Comparison with distances estimated from stellar parameters}

Second test was done using the comparison of the distances estimated from {\em WISE} photometry and 
spectrophotometric parallaxes. \citet{Saguner11} derived atmospheric parameters of 245 faint and high 
Galactic latitude RC stars from high signal-to-noise middle resolution spectra and calculated their 
spectrophotometric distances. We removed 23 of 245 stars from the sample as their distance errors are 
too high, decreasing the number of stars to 222. These RC stars with $200 \lesssim d \lesssim 500$ pc 
give us the chance to compare their distances estimated by two different, photometric and 
spectrophotometric, methods.

We obtained the {\em WISE} magnitudes of the stars from the {\em WISE} All-Sky Data Release 
\citep{Cutri12} according to the equatorial coordinates in \citet{Saguner11}. The {\em WISE} magnitudes 
and colour of these 222 stars were de-reddened using the colour excesses in \citet{Saguner11} 
and Eq. 2. in this study. Distances of the RC stars were evaluated by Pogson's equation using the 
absolute magnitudes in Table 1. The distances calculated using the adopted absolute magnitudes
$M_{W1} = -1.635\pm 0.026$ and $M_{W3} = -1.606\pm 0.024$ mag of the RC stars with 
$\sigma_\pi / \pi \leq 0.15$ are compared with the ones in \citet{Saguner11} in Fig. 11a and c, 
respectively. Residuals from 1-1 line are also shown in Fig. 11b and d. These figures demonstrate 
that there is good agreement between the distances calculated using the both methods. The mean of 
the differences between two sets of distances and the corresponding standard deviations are 
$<\Delta d>=-9.2$ pc and $\sigma=30.8$ pc for Fig 11a and $<\Delta d>=-9.6$ pc and $\sigma=30.7$ pc 
for Fig 11c. There are 153 stars in both panels within the $\pm1\sigma$ limits, 69 per cent of the 
sample.

\section{Conclusions}

In this study, we identified the RC stars with the {\em WISE} photometry for the first time. We also 
identified them with the 2MASS photometry, in order to compare their {\em WISE} absolute magnitudes 
in this study with the ones evaluated from the transformation equations \citep{Bilir12}. Another 
comparison was made between the distances obtained using the {\em WISE} absolute magnitudes and those 
taken from \citet{Saguner11} who used a spectrophotometric method. 

Conclusions of this study can be summarized as follows,

\begin{itemize}

\item The RC stars occupy a very small region in the {\em WISE} colour-magnitude diagram of the 
stars in the Solar neighbourhood, as expected from the general properties of these stars in other 
photometric systems. 

\item The adopted {\em WISE} absolute magnitudes and colour of the RC stars are 
$M_{W1}=-1.635\pm 0.026$, $M_{W3}=-1.606\pm 0.024$ and $(W1-W3)_0=-0.028\pm 0.001$ mag.

\item As for the effect of LK-corrected parallaxes on the results, it is found that selecting different 
limits for the relative parallax errors, provided that it is higher than 0.10, does not considerably 
affect the mean absolute magnitudes and the colour of the RC stars. 

\item The 2MASS absolute magnitudes and colours of the RC stars are $M_{J}=-0.970\pm 0.016$, 
$M_{H}=-1.462\pm 0.014$, $M_{K_{s}}=-1.595\pm 0.025$, $(J-H)_0=0.485\pm 0.002$, 
$(H-K_{s})_0=0.130\pm 0.002$ and $(J-K_{s})_0=0.612\pm 0.003$ mag.

\item The {\em WISE} absolute magnitudes and colour of the RC stars are in good agreement with the 
semi-empirical ones evaluated from the transformation equations.

\item From the comparisons of the distances estimated using the {\em WISE} absolute magnitudes and 
the ones calculated using a spectrophotometric method, we conclude that agreement of two sets of 
distances confirms the confidence of the absolute magnitudes in question. 

\end{itemize}

As stated above, {\em WISE} photometry goes one magnitude deeper than the 2MASS photometry for 
sources with spectra close to that of an A0 star, and even deeper for moderately red 
sources like K stars or galaxies with old stellar populations. Hence, we suggest that the absolute 
magnitudes estimated in this study can be used in the distance calculations for new sets of the 
RC stars at relatively larger distances. For example, these values can be used to estimate the 
Galactic model parameters, in order to test any degeneracy between different parameters of the 
Galactic components.

\section{Acknowledgments}

This work has been supported in part by the Scientific and Technological Research 
Council (T\"UB\.ITAK) 112T120. We would like to thank the anonymous referee for comments and 
suggestions. This research has made use of the Wide-field Infrared Survey 
Explorer and NASA/IPAC Infrared Science Archive and Extragalactic Database (NED), which are 
operated by the Jet Propulsion Laboratory, California Institute of Technology, under contract 
with the National Aeronautics and Space Administration. This publication makes use of data products 
from the Two Micron All Sky Survey, which is a joint project of the University of Massachusetts and 
the Infrared Processing and Analysis Center/California Institute of Technology, funded by the 
National Aeronautics and Space Administration and the National Science Foundation. This research 
has made use of the SIMBAD, and NASA's Astrophysics Data System Bibliographic Services.

\pagebreak[4]

\begin{figure*}
\begin{center}
\includegraphics[scale=0.30, angle=0]{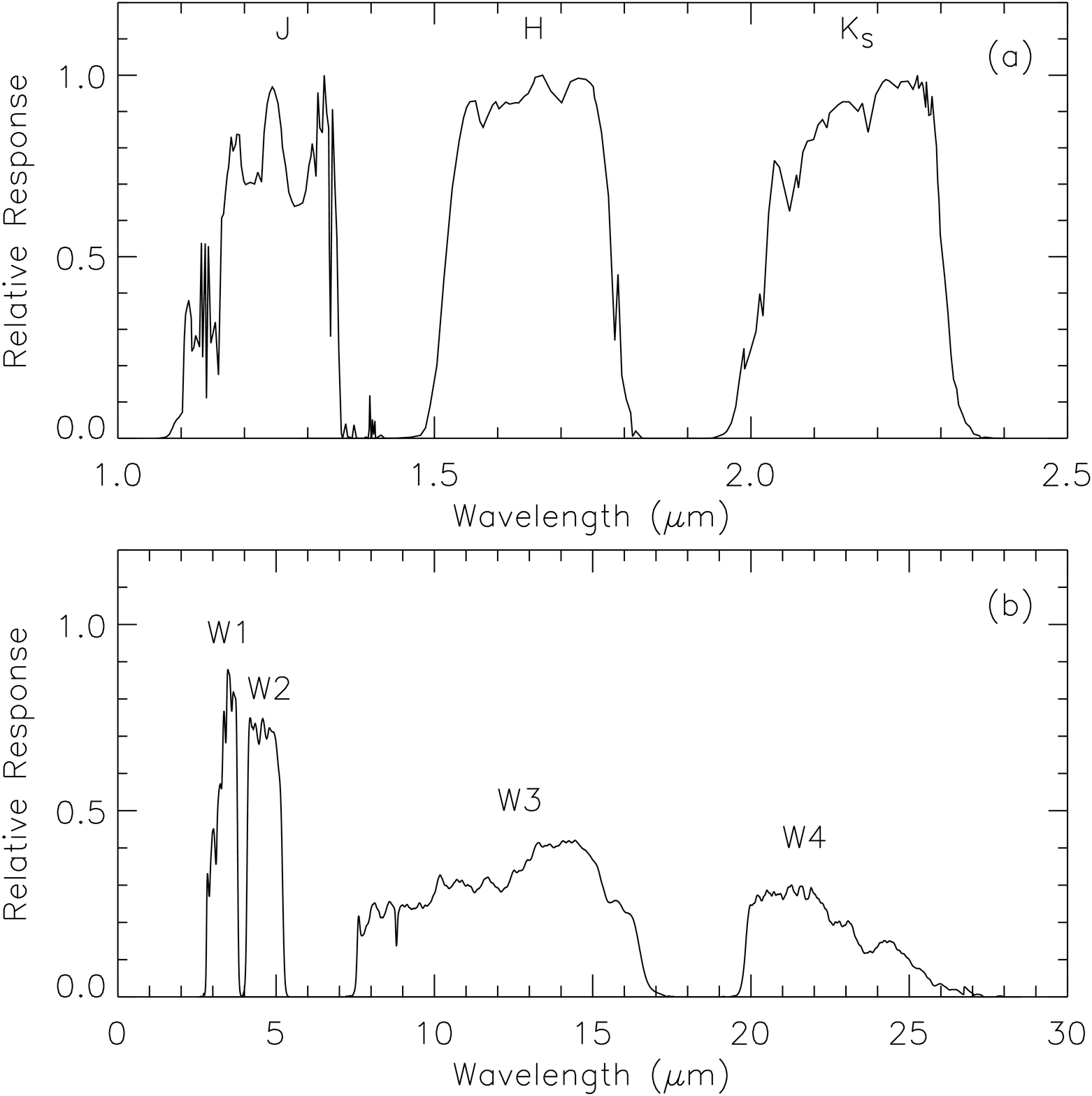}
\caption[]{Relative responses of the 2MASS (a) and {\em WISE} filters (b).}
\label{Fig1}
\end{center}
\end{figure*}

\begin{figure*}
\begin{center}
\includegraphics[scale=0.20, angle=0]{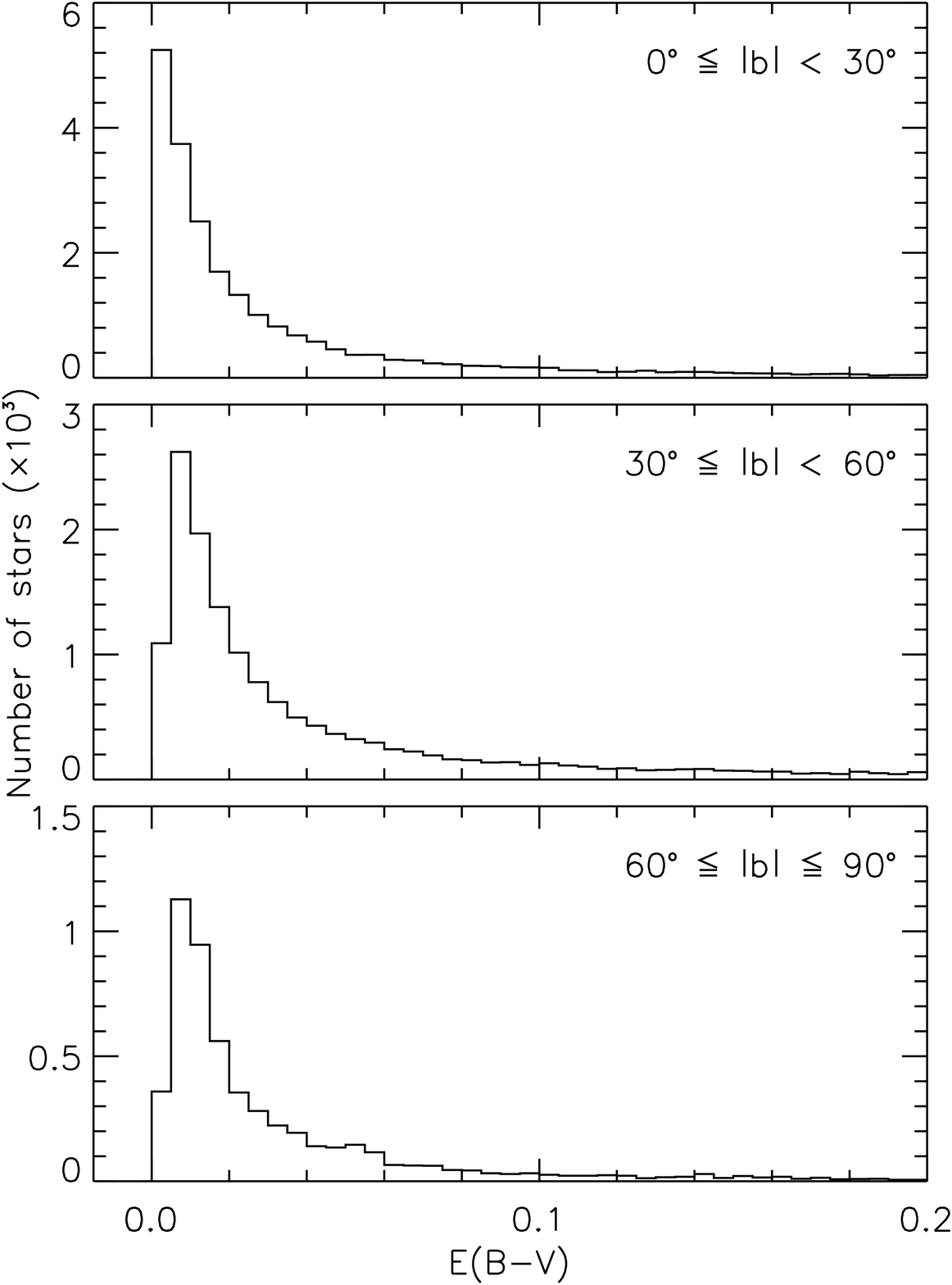}
\caption[]{Distributions of the reduced colour excesses of the 46,214 stars 
in our sample according to the Galactic latitude ($b$).} 
\label{Fig2}
\end{center}
\end{figure*}

\begin{figure*}
\begin{center}
\includegraphics[scale=0.40, angle=0]{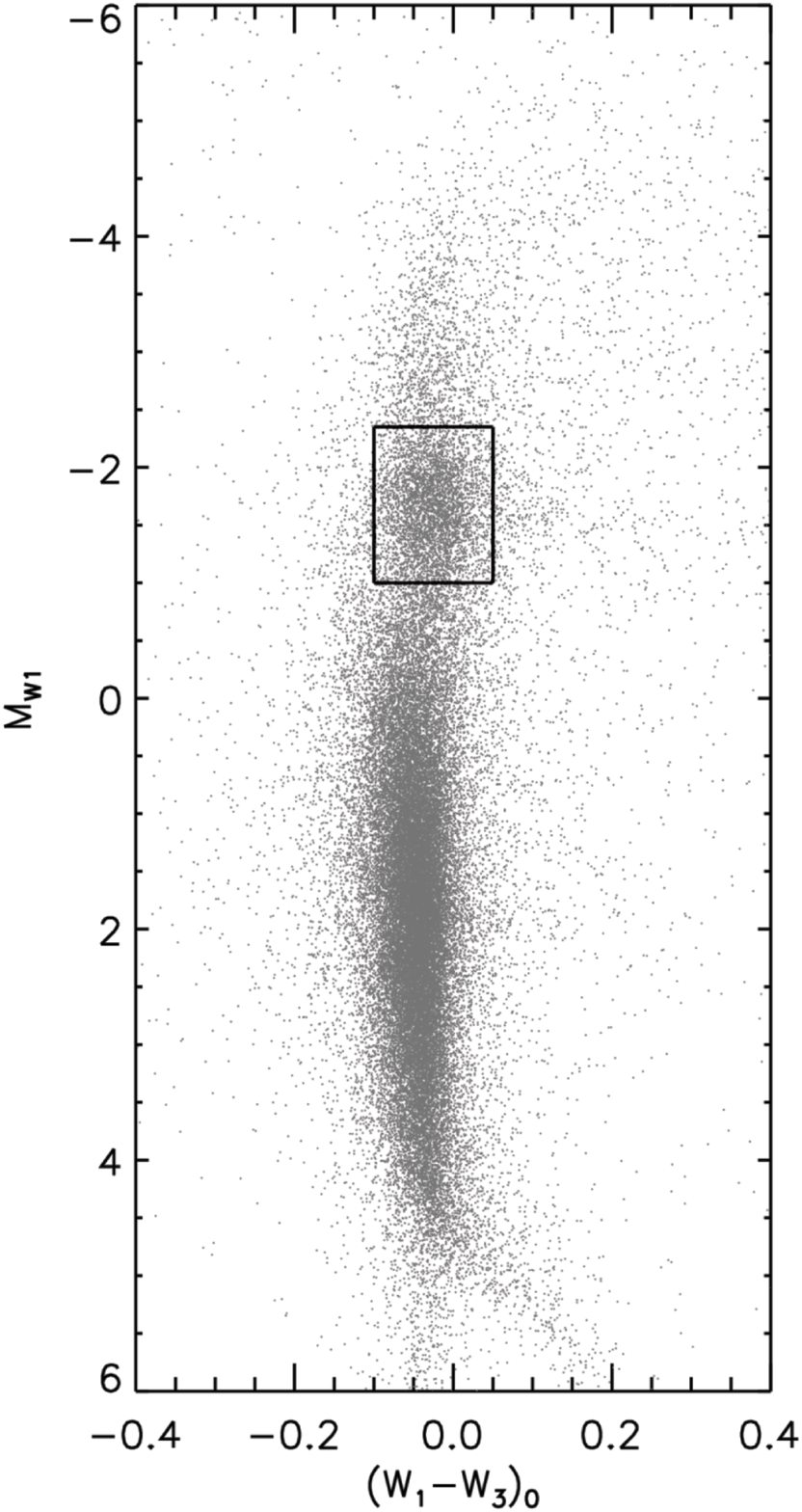}
\caption[]{The $M_{W1}$ absolute magnitudes versus $(W1-W3)_0$ colours for 46,214 
stars with the relative parallax errors $\sigma_\pi / \pi \leq 0.15$ selected 
from the {\em Hipparcos} catalogue. Stars in the rectangle are adopted as 
the RC stars.} 
\label{Fig3}
\end{center}
\end{figure*}

\begin{figure*}
\begin{center}
\includegraphics[scale=0.20, angle=0]{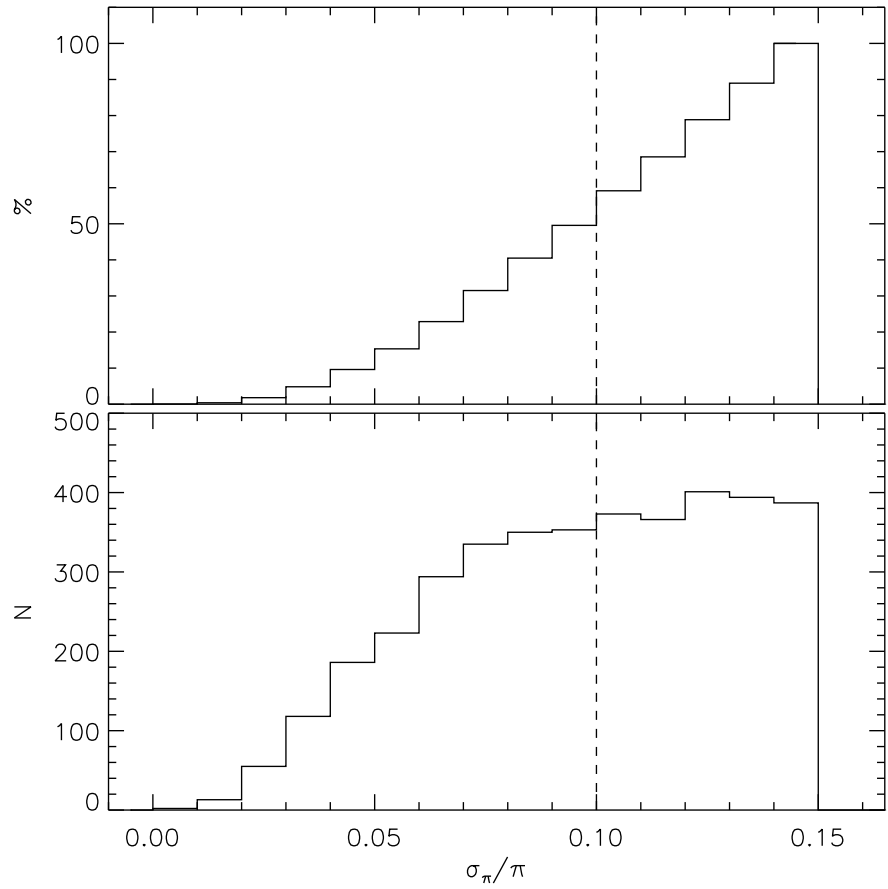}
\caption[]{Number (lower panel) and cumulative (upper panel) distributions of the 
relative parallax errors for the 3889 RC stars selected from {\em WISE} photometry. 
Dashed lines represent the median value.} 
\label{Fig4}
\end{center}
\end{figure*}

\begin{figure*}
\begin{center}
\includegraphics[scale=0.30, angle=0]{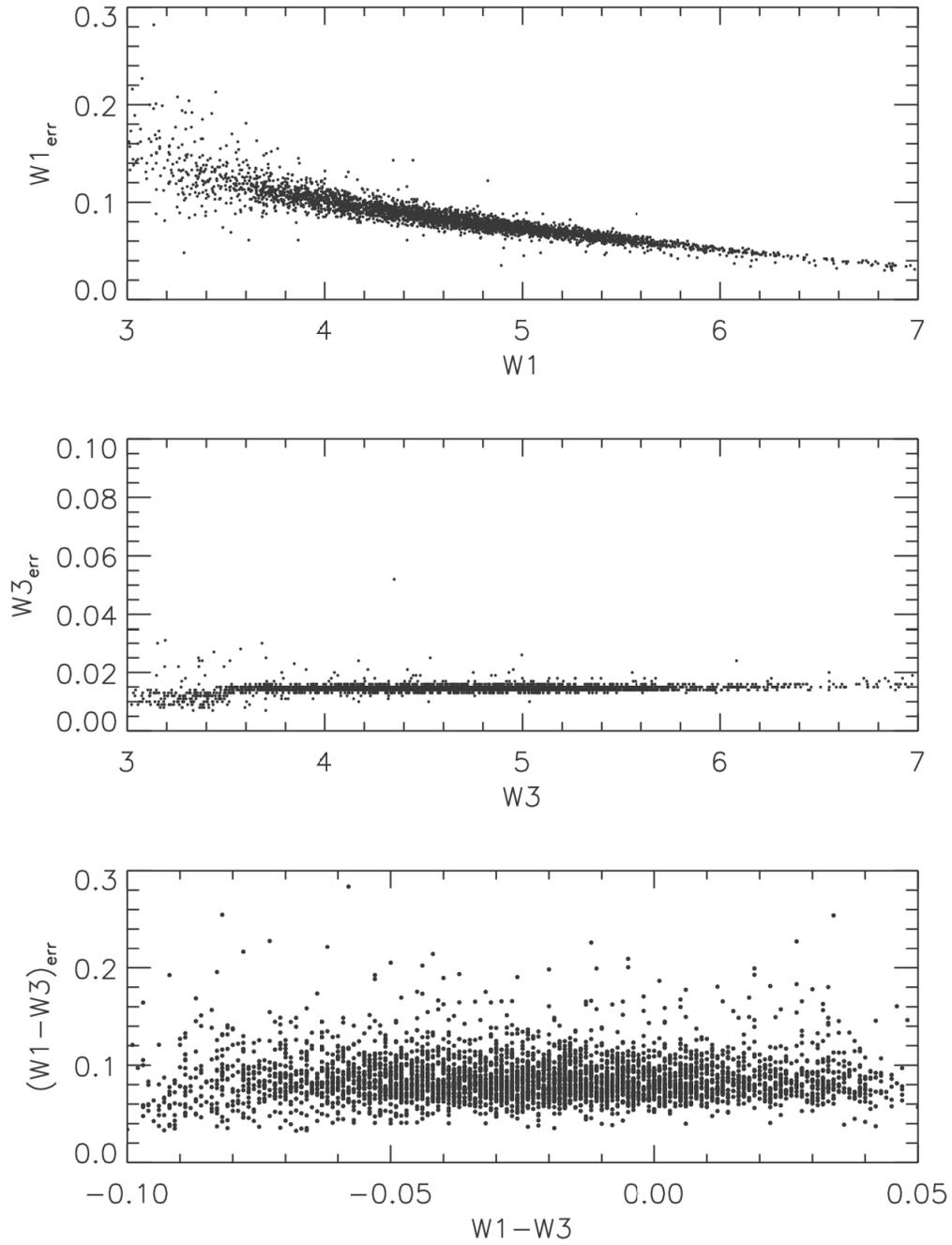}
\caption[]{Photometric errors of the 3889 RC stars selected from the {\em WISE} data.} 
\label{Fig5}
\end{center}
\end{figure*}

\begin{figure*}
\begin{center}
\includegraphics[scale=0.50, angle=90]{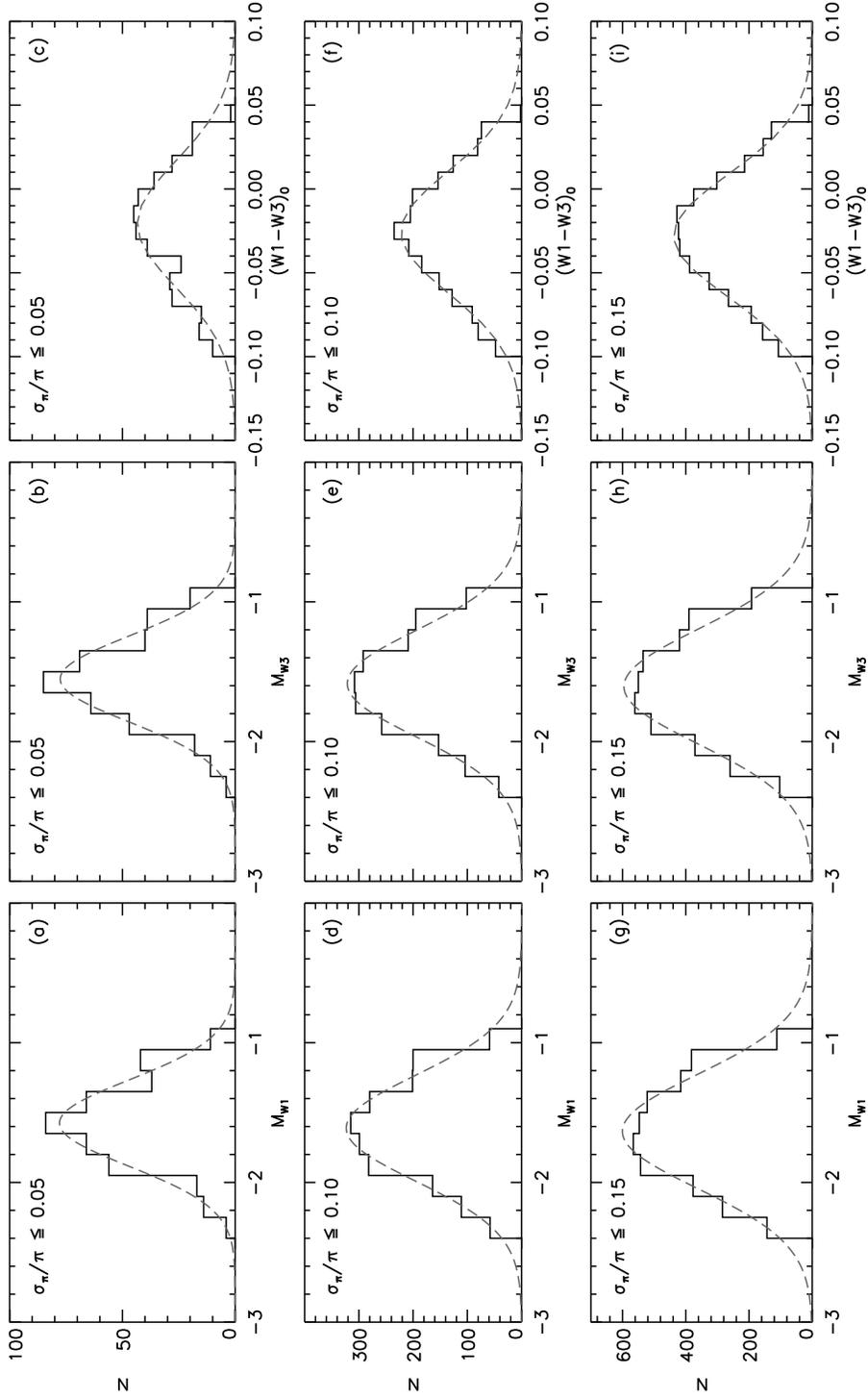}
\caption[]{Histograms of $M_{W1}$ and $M_{W3}$ absolute magnitudes and $(W1-W3)_0$ colour 
for the RC stars with relative parallax errors $\sigma_\pi / \pi \leq 0.05$ (a-c), 
$\sigma_\pi / \pi \leq 0.10$ (d-f) and $\sigma_\pi / \pi \leq 0.15$ (g-i). Gaussian functions 
fitted to the distributions are shown with dashed lines.}
\label{Fig6}
\end{center}
\end{figure*}

\begin{figure*}
\begin{center}
\includegraphics[scale=0.35, angle=0]{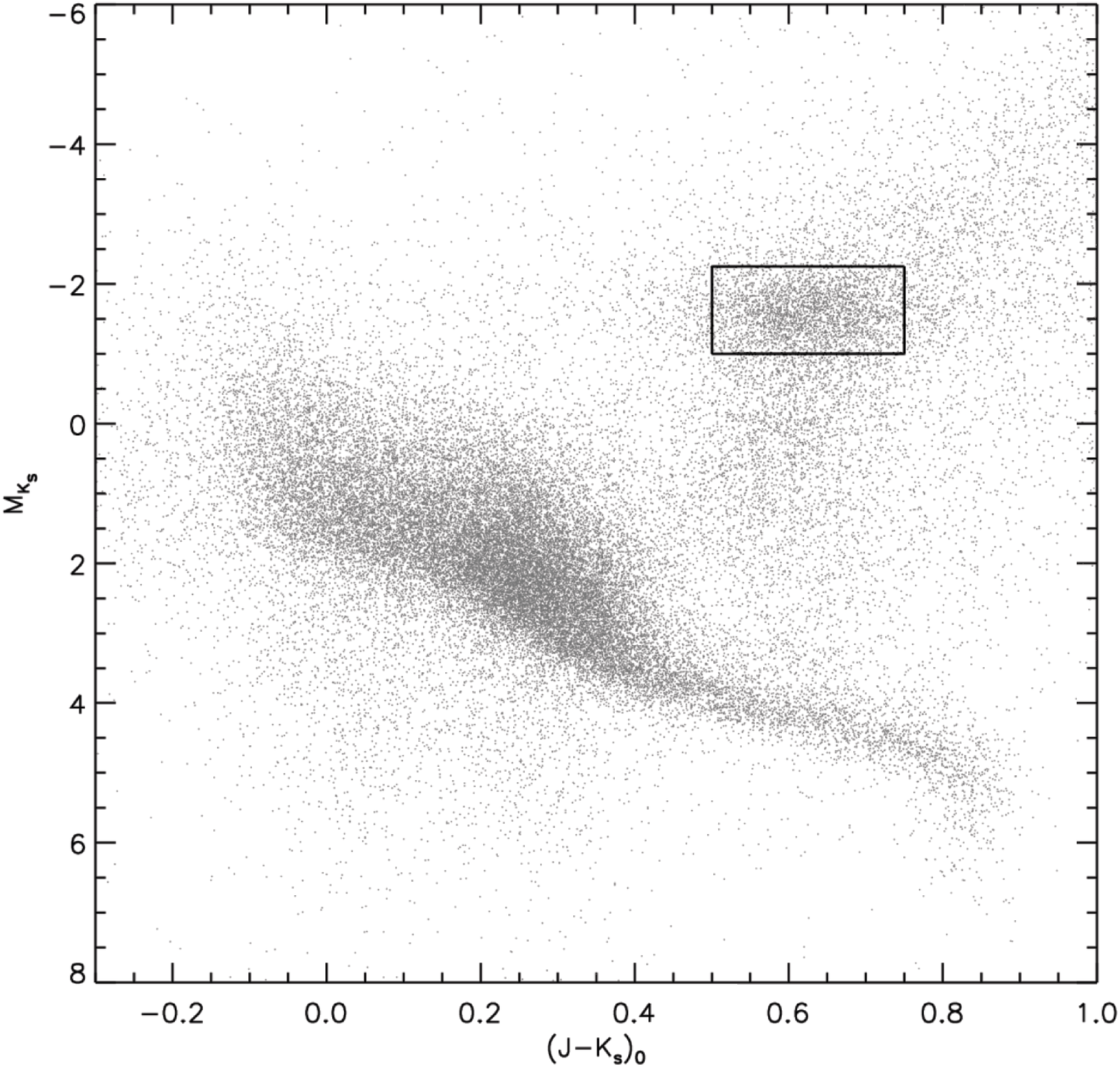}
\caption[]{The $M_{K_{s}}$ absolute magnitudes versus $(J-K_{s})_0$ colours for 
46,214 stars with the relative parallax errors $\sigma_\pi / \pi \leq 0.15$ 
selected from the {\em Hipparcos} catalogue. Stars in the rectangle are adopted as 
the RC stars.} 
\label{Fig7}
\end{center}
\end{figure*}

\begin{figure*}
\begin{center}
\includegraphics[scale=0.20, angle=0]{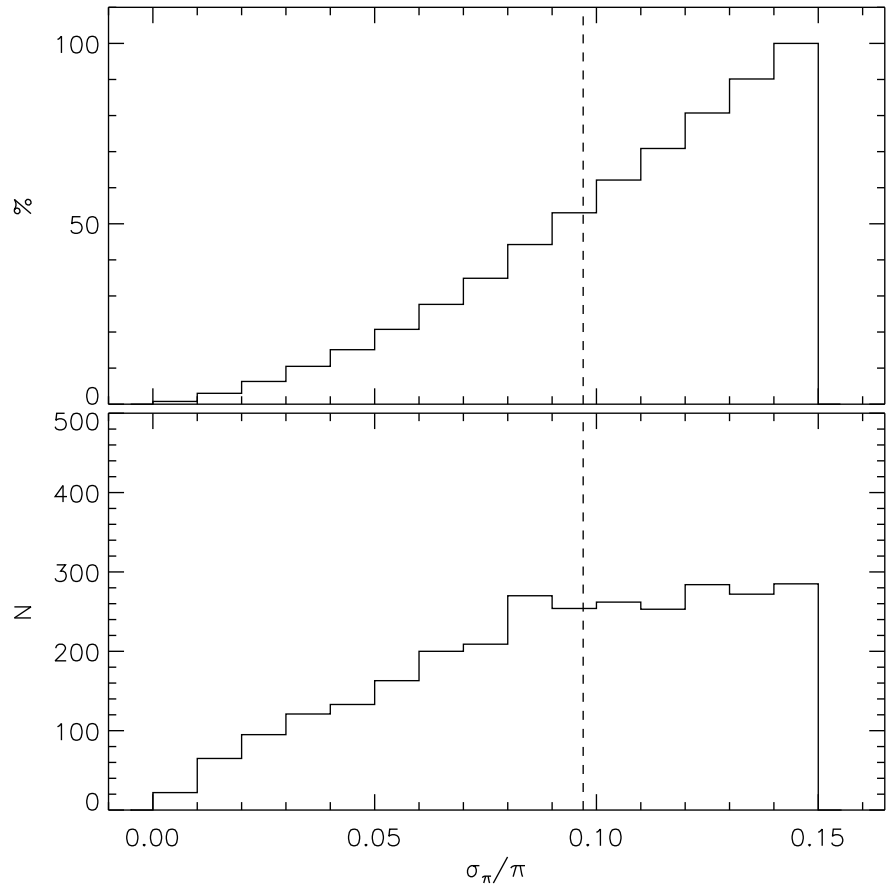}
\caption[]{Number (lower panel) and cumulative (upper panel) distributions of 
the relative parallax errors for the 2937 RC stars in the 2MASS photometry. 
Dashed lines represent the median value.} 
\label{Fig8}
\end{center}
\end{figure*}

\begin{figure*}
\begin{center}
\includegraphics[scale=0.25, angle=0]{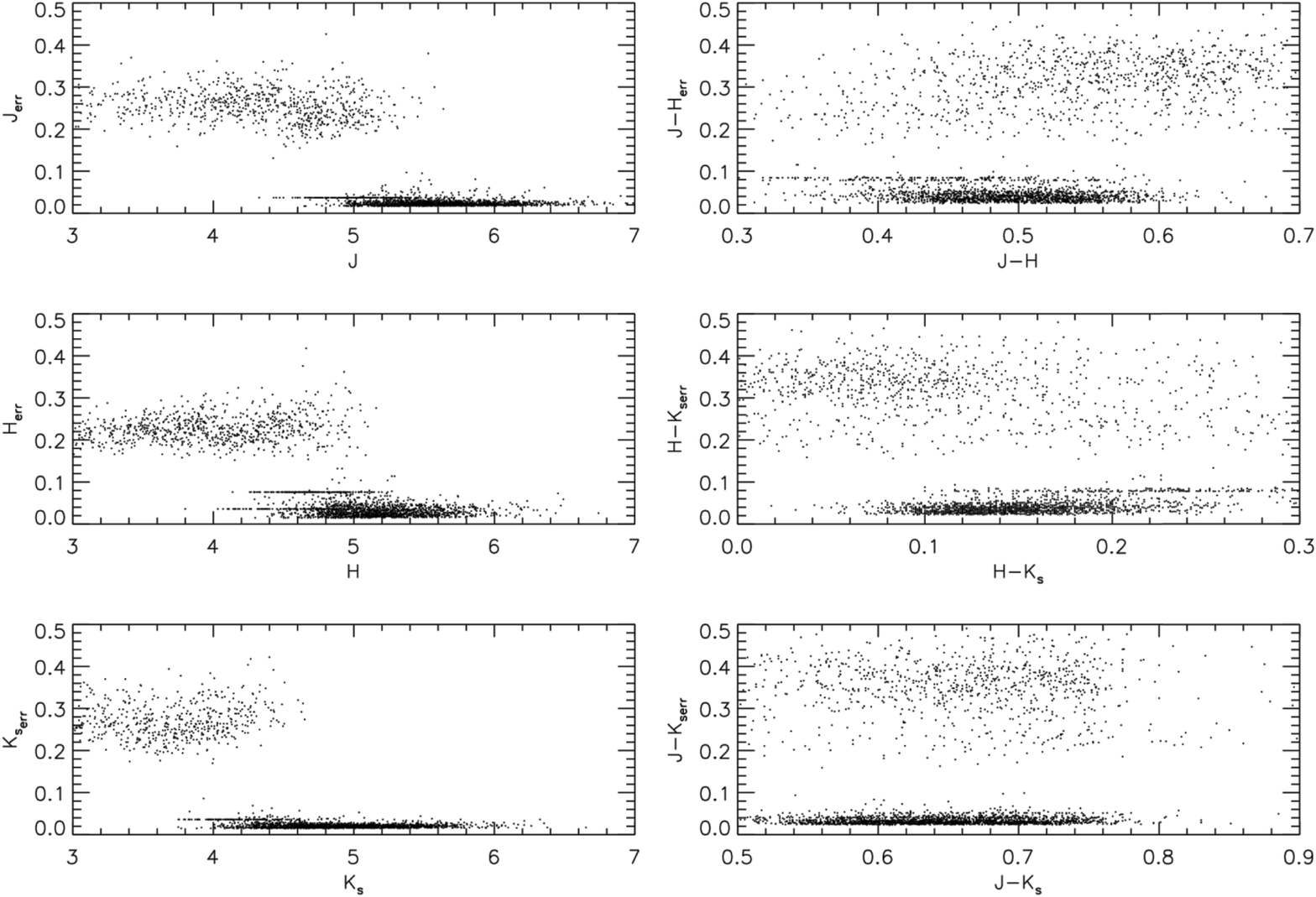}
\caption[]{Photometric errors of the 2MASS observations for the 2937 RC stars 
selected from the $M_{K_{s}}-(J-{K_{s}})$ colour-magnitude diagram.} 
\label{Fig9}
\end{center}
\end{figure*}

\begin{figure*}
\begin{center}
\includegraphics[scale=0.60, angle=0]{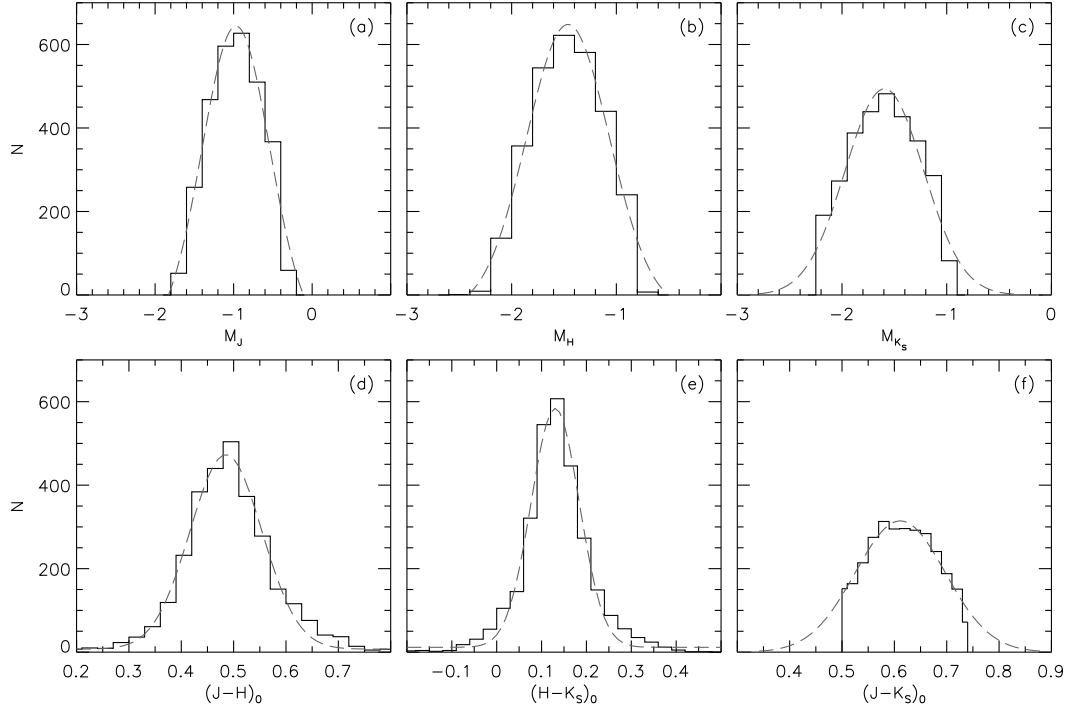}
\caption[]{Distributions of the $M_{J}$, $M_{H}$ and $M_{K_{s}}$ absolute 
magnitudes (a-c) and $(J-H)_0$, $(H-K_{s})_0$ and $(J-K_{s})_0$ colours (d-f) 
for the 2937 RC stars in our sample. Gaussian fits to the distributions are 
also shown.} 
\label{Fig9}
\end{center}
\end{figure*}

\begin{figure*}
\begin{center}
\includegraphics[scale=0.4, angle=0]{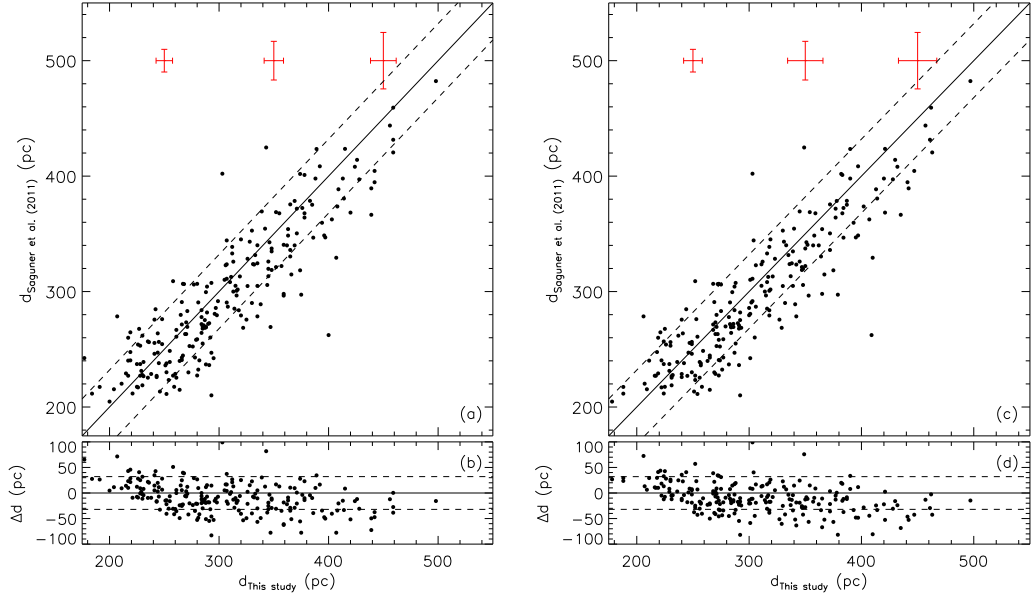}
\caption[]{Comparison of the distances for the 222 RC stars estimated in 
\citet{Saguner11} and in this study using the adopted $M_{W1}$ (a) and $M_{W3}$ (c). 
Distributions of their differences relative to the distances estimated in our 
study (b and d). Solid lines represent the 1-1 values, dashed lines 
$\pm 1\sigma$ limits.} 
\label{Fig9}
\end{center}
\end{figure*}

\end{document}